\documentclass[pre,twocolumn,showpacs]{revtex4}

\usepackage[dvips]{graphicx}

\usepackage{amsmath}

\DeclareMathOperator{\erf}{erf}

\DeclareMathOperator{\erfc}{erfc}

\newcommand{\D}{\ensuremath{\mathrm{d}}}

\newcommand{\rme}{\ensuremath{\mathrm{e}}}

\newcommand{\abreact}{$A + B \rightarrow B$ }

\newcommand{\bcreact}{$B + C \rightarrow C$ }

\newcommand{\bothreact}{$A + B \rightarrow B$, $B + C \rightarrow C$ }

\begin{document}


\title{Survival probabilities in the double trapping reaction 
\bothreact}

\date{\today}

\author{Alan J. Bray}

\author{Richard Smith}

\affiliation{School of Physics and Astronomy, University of 
Manchester, Manchester M13 9PL, U.K.}

\begin{abstract}
We consider the double trapping reaction \bothreact in one dimension. 
The survival probability of a given $A$ particle is calculated under 
various conditions on the diffusion constants of the reactants, and 
on the ratio of initial B and C particle densities. The results are shown 
to be of more general form than those obtained in previous work on the
problem.

\end{abstract}


\maketitle

\section{Introduction}

The simple trapping reaction  \abreact has been studied intensely over
the  last two decades,  motivated in  part by  studies of  the related
two-species annihilation  problem, $A  + B \to  0$, introduced  in the
classic  paper of  Toussaint  and Wilczek  \cite{TW}.  In one  spatial
dimension the  survival probability,  $Q_A(t)$, of a  single diffusing
$A$-particle moving in an intially homogeneous background of diffusing
$B$  particles has  been proved  rigorously to  decay as  $Q_A(t) \sim
\exp(-\lambda  t^{1/2})$ \cite{BL}.   Only recently,  however,  has an
exact   expression  for   the  coefficient   $\lambda$   been  derived
\cite{BB1}:  $\lambda   =  (4/\sqrt{\pi})\,\rho_B  \sqrt{D_B}$,  where
$\rho_B$ and $D_B$ are the density and diffusion constant respectively
of the  $B$ particles. The  result has been generalized  \cite{BB2} to
all (continuous) dimensions $d \le 2$. A curious feature of the result
is the absence of any  dependence on the diffusion constant, $D_A$, of
the $A$  particle (at least in  the leading asymptotics  -- $D_A$ does
appear,  however,  in  subdominant  terms \cite{AB}).  For  $d>2$  the
rate-equation approach, which predicts  a simple exponential decay, is
qualitatively correct \cite{BL}.

In  contrast to the  simple trapping  reaction, more  complex trapping
sequences have received little attention. This paper is devoted to the
double trapping  reaction \bothreact.  We try to  compute the fraction
of  $A$  particles  remaining   at  time  $t$,  or  equivalently,  the
probability, $Q_A(t)$, that  a single $A$ particle has  survived up to
time $t$,  given that it is  initially surrounded by a  sea of Poisson
distributed traps, the $B$ particles. The $B$ particles themselves are
initially surrounded by Poisson distributed traps -- the $C$ particles
-- and  thus  disappear from  the  system at  each  time  step with  a
probability to be determined.  By ``Poisson distributed'' we mean that
the probability  to find  $N$ traps  in an interval  of length  $L$ is
$[(\rho_{\mathrm{B,C}} L)^N/N!]  \exp (-\rho_{\mathrm{B,C}} L)$, where
$\rho_{\mathrm{B,C}}$ is  the density of $B$ or  $C$ particles.  Since
the $C$ particles diffuse independently, and do not disappear from the
system, they remain  Poisson distributed at  all times.  The spatial
distribution  of  the  $B$  particles  may  change,  however,  as  $B$
particles are absorbed.

Of  particular  interest is  the  limit  $t  \to \infty$.   Since  the
$B$-particle density ultimately decays  to zero, the function $Q_A(t)$
will  approach a  non-zero limit  $Q_A(\infty$) which,  on dimensional
grounds,  can only  depend  on the  density ratio  $\rho_{B0}/\rho_C$,
where $\rho_{B0} = \rho_B(0)$ is the initial $B$-particle density, and
on  the  ratios  $D_A/D_B$,  $D_B/D_C$ of  diffusion  constants.   The
calculation of $Q_A(\infty)$ is our main goal.

This double trapping problem has  been studied using a mean-field (i.e.
rate equation) approach, which should again be qualitatively valid for
$d>2$, and by a version of the Galanin model for $d=1$ \cite{sanchez}.
The latter approach predicts the following limiting result:
\begin{equation} 
\label{sanchez}
Q_A(\infty)  = \left(1 + \frac{\rho_{B0}}{\rho_C}
\sqrt{\frac{D_A+D_B}{D_B+D_C}} \right)^{-1}.
\end{equation}
This paper  considers various limiting regimes of  the general problem
in which exact results or bounds can obtained for comparison with  Eq.\
(\ref{sanchez}).  We  also obtain some results on  the asymptotic form
of the time-dependence  of $Q_A(t)$, i.e.\ on the  manner in which the
asymptotic limit is approached.

\section{Analysis of the survival probabilities}

Each  particle  $i$  diffuses   according  to  the  Langevin  equation
$\dot{x_i} = \eta_i  (t)$, where $\eta_i (t)$ is  Gaussian white noise
with mean zero and correlator  $\langle \eta_i (t) \eta_j (t') \rangle
= 2  D \delta_{ij} \delta (t-t')$,  where $D = D_A$,  $D_B$, $D_C$ are
the diffusion constants for the  three reactant species. From this one
can  derive  in standard  fashion  a  backward Fokker-Planck  equation
governing the  time evolution  of the particle's  survival probability
$Q(x_i,t)$:
\begin{equation}
\frac{\partial    Q(x_i,t)}{\partial   t}    =    D   \frac{\partial^2
Q(x_i,t)}{\partial x_i^2}, \label{fokker}
\end{equation}
where $x_i$ is the position of the ith particle at time $t = 0$.

We consider three subsets of  the general double trapping problem: the
cases where $D_A  \ll D_B \ll D_C$, where $D_B =0$ and where $D_A =  
D_C =0$.  In
the first two cases it will  be necessary, to obtain exact results or 
exact bounds, to limit consideration to the regime $\rho_{B0} \ll \rho_C$. 
For the case $D_A = D_C =0$, however, an exact result for $Q_A(\infty)$ 
is possible for all values of the ratio $\rho_{B0}/\rho_C$.

\subsection{The case $\bf{D_A \ll D_B \ll D_C}$} 

We first consider a subset  of the general \bothreact problem in which
the diffusion constants of the  $A$, $B$ and $C$ particles are subject
to the condition $D_A \ll D_B \ll D_C$.  This allows us to approximate
each  process  as   an  independent  ``target  annihilation  problem''
\cite{blumen},  i.e.\  in  the  reaction  \bcreact we  treat  the  $B$
particles as static and the $C$  particles as mobile traps, and in the
\abreact process we  consider the $B$ particles as  mobile traps for a
static $A$  particle.  We also  subject this problem to  the condition
$\rho_{B0} \ll \rho_C$ to ensure that the $B$ particles remain Poisson
distributed  even at  large  times,  i.e. that  no  clustering of  $B$
particles emerges  in regions  free of $C$  particles. The  reason for
this condition will become clear shortly.

We begin by considering a single, static $B$ particle at the origin. 
The survival probability, $Q_1$, of the $B$ particle with only one 
$C$ present, starting at some $x>0$, is the solution of (\ref{fokker}) 
subject to the boundary conditions $Q_1(0,t) = 0$, $Q_1(\infty,t)=1$ 
and $Q_1(x,0)=1$. That is~\cite{redner}
\begin{equation*}
Q_1(x,t) = \erf \left( \frac{x}{\sqrt{4 D_c t}} \right), \label{erf}
\end{equation*}
where $\erf(y)$ is the error function. With $N$ traps present,
where $N=\rho_C L$, starting at positions $x_i$ uniformly distributed 
in $[0,L]$, the survival probability of the $B$ particle is (following 
the argument in~\cite{blumen})
\begin{equation*}
Q_N(t)= \prod_{i=1}^N \left[ \frac{1}{L} \int_0^L \D x_i 
\erf \left( \frac{x_i}{\sqrt{4 D_c t}}\right) \right],
\end{equation*}
where we have averaged over the starting positions. Rewriting the error 
function in terms of the complementary error function, $\erf(y) = 1 - 
\erfc(y)$, and using the fact that that the C particles are independent, 
we have
\begin{equation}
Q_N(t) = \left[ 1 - \frac{1}{L} \int_0^L \D x \erfc \left(
\frac{x}{\sqrt{4 D_c t}} \right) \right]^{\rho_c L}. \label{qN}
\end{equation}
Taking the limits $N \to \infty$ and $L \to \infty$, keeping
$\rho_C$ fixed, and evaluating the integral in (\ref{qN}), gives
\begin{equation}
Q(t) = \exp \left( - \frac{2}{\sqrt{\pi}}\,\rho_C \sqrt{D_C t} \right). 
\label{oneside}
\end{equation}

We can perform the same calculation on the opposite side of the $B$
particle and, since the results are symmetric and independent, get
the same result. So the full survival probability of the $B$
particle is simply the square of (\ref{oneside}), i.e.
\begin{equation}
Q_B(t) = \exp \left( - \frac{4}{\sqrt{\pi}}\,\rho_C \sqrt{D_C t} \right),
\label{qB}
\end{equation}
which is the standard result for the one-dimensional target problem.

For the \abreact process we can use the `toy model' introduced 
in~\cite{odon}. In this model, the traps $B$ are assumed to disappear 
randomly in a manner consistent with the required density $\rho_B(t)$. 
The model will be an exact representation of the double trapping reaction 
provided $\rho_{B0} \ll \rho_C$, so that no correlations develop in the 
positions of surviving $B$-particles. The time-dependence of the 
$A$-particle survival probability, $Q_A(t)$, within this model is 
calculated  using a similar argument to that outlined above, 
but with the traps disappearing from the system at each time step with a 
known probability. In our case, we can describe this decay of traps using 
the survival probability of a $B$ particle, $Q_B(t)$, as given by~(\ref{qB}). 
The model gives for the survival probability of the $A$ particle~\cite{odon} 
\begin{equation*}
Q_A(t) = \exp \left( -2 \rho_{B0} \sqrt{\frac{D_B}{\pi}} 
\int_0^t \frac{\D \tau}{\tau^{1/2}} Q_B(\tau) \right). 
\label{toymodel}
\end{equation*}
Substituting for $Q_B(\tau)$ from Eq.\ (\ref{qB}) gives the result
\begin{equation}\label{qA}
Q_A(t) = \exp \left[ - \frac{\rho_{B0}}{\rho_C} \sqrt{\frac{D_B}{D_C}} 
\left( 1 - \rme^{-\frac{4}{\sqrt{\pi}}\rho_C \sqrt{D_C t}} \right) \right].
\end{equation}
Note that this result was derived treating the $A$ particles as static, so 
it will become asymptotically exact in the limit $D_A \ll D_B$ (and we have 
already assumed $D_B \ll D_C$). 

For  the case where  $\rho_{B0} \ll  \rho_C$ and  $D_B \ll  D_C$ still
hold, but $D_A$ is  arbitrary, the ``Pascal principle'' \cite{Pascal},
according to  which the $A$ particle  survives longest if  it does not
move, shows that  Eq.\ (\ref{qA}) provides an upper  bound on $Q_A(t)$
for any value of $D_A$.

\subsection{The case ${\bf D_B=0}$} 
\begin{figure}
\includegraphics[width=\linewidth]{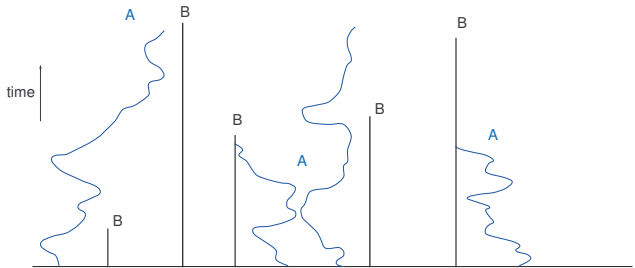}
\caption{\label{dbzero} For $D_B = 0$ and $\rho_{B0} \ll \rho_C$, the $A$ 
particles diffuse among static traps which disappear randomly and 
independently.}
\end{figure}

For the \bcreact  reaction, with the B particles  remaining static, we
can  again use  the result  in  Eq.\ (\ref{qB})  for the  $B$-particle
survival  probability.  The   \abreact  process  remains,  however,  a
nontrivial problem, consisting of $A$ particles diffusing  among static
traps each of  which has a survival probability  $Q_B(t)$, as shown in
fig.~\ref{dbzero}.  Taking  once   again  the  limit  $\rho_{B0}  \ll\
\rho_C$, the survival probabilities of the different $B$-particles can
be treated as independent, but the problem is still nontrivial.
We  can, however, find  a lower  bound on  $Q_A(t)$ by  considering an
approach based on  the span, $R(t)$, of a random  walk -- the distance
from  the point  of  maximum  excursion in  one  direction to  maximum
excursion in the opposite direction, up to time $t$. The average value
of this quantity is given by (see ~\cite{weiss})
\begin{equation}\label{R}
\langle R(t) \rangle = 4\sqrt{\frac{D_A t}{\pi}}.
\end{equation}
We formulate the \abreact process as follows. We write the
infinitesimal change in the survival probability of a given $A$ particle,
$dQ_A$, in terms of the probability that it is trapped by a $B$
particle in the time interval $t \to t+dt$ in which the span of
the random walk increases by $dR$:
\begin{eqnarray}\label{spanarg}
\nonumber dQ_A &=& - \rho_{B0}\; dR \; Q(A,B;t) \\
&=& -\rho_{B0}\; dR\; Q_A\; Q(B|A;t),
\end{eqnarray}
where  $\rho_{B0}  dR$ is  the  probability  to  find a  $B$  particle
initially  in  the interval  $dR$  (treating,  as  usual, the  initial
$B$-particle locations as a  Poisson process), $Q(A,B;t)$ is the joint
probability  that both  the $A$  particle  and the  $B$ particle  have
survived up to time $t$, and $Q(B|A;t)$ is the conditional probability
that the $B$  particle has survived until time $t$  given that the $A$
particle has survived. To treat the survival probability of the $A$
and the $B$ particles as independent, we must make the assumption that
the $B$ particles  remain Poisson distributed at all  times, i.e. that
the positions of the $B$  particles are not spatially correlated. If the
$B$ particles  survive in clusters  and there develop  regions totally
free of $B$  particles, then the probability that  an $A$ particle has
survived will  depend on whether  it is in  a $B$-free region or  in a
cluster  of $B$  particles. In the  limit $\rho_{B0}  \ll \rho_C$, the 
assumption of independence is justified, and allows us to write  
$Q(B|A;t)=Q_B(t)$,  with  $Q_B(t)$  given  by  (\ref{qB}).  The
solution of (\ref{spanarg}) is then
\begin{equation*} Q_A(t) = \exp \left( - \rho_{B0} \int_0^t
Q_B(\tau) \; \frac{dR}{d\tau}\; d\tau \right).
\end{equation*}

The  span is  a stochastic  variable so  we need  to average  over all
realizations of the function $R(\tau)$, which analytically is not trivial.  
We can, however, obtain a lower bound by using the convexity inequality
\begin{eqnarray}\label{convex}
\nonumber \langle Q_A(t) \rangle &=& \left< \exp \left( - \rho_{B0}
\int_0^t Q_B(\tau) \; \frac{dR}{d\tau}\; d\tau \right) \right> \\
\nonumber &\geq & \exp \left( - \rho_{B0} \int_0^t Q_B(\tau) \; 
\frac{d}{d\tau} \langle R \rangle\; d\tau \right),
\end{eqnarray}
and by substituting for $\langle R(t) \rangle$ using (\ref{R}) we
obtain the result
\begin{equation}\label{lowbound}
\langle Q_A(t) \rangle \geq \exp \left[ -\frac{\rho_{B0}}{\rho_C}
\sqrt{\frac{D_A}{D_C}} \left( 1- e^{-\frac{4}{\sqrt{\pi}} \rho_C
\sqrt{D_C t}} \right) \right].
\end{equation}

\subsection{The case ${\bf D_A = D_C = 0}$} 
\begin{figure}
\includegraphics[width=\linewidth]{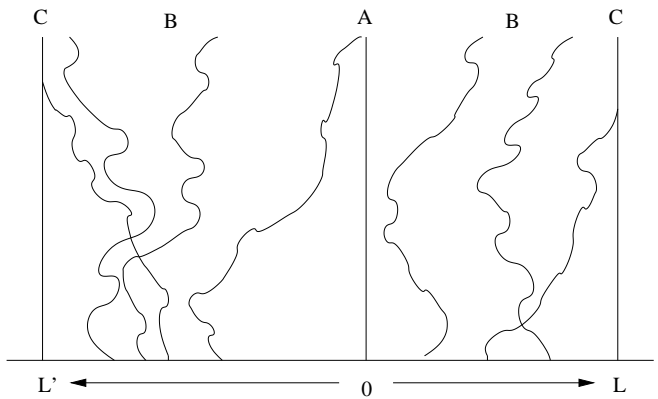}
\caption{\label{daczero} We consider the subset of the problem where 
$D_A = D_C = 0$. For the $A$ particle to survive, all the $B$ particles 
between the neighbouring $C$ particles must reach a $C$ particle first, 
thus being removed from the system before reaching the $A$ particle.}
\end{figure}

We now consider the case where the $A$ and $C$ 
particles remain static and the $B$ particles diffuse among them. We can 
treat this is an extension of the Gambler's Ruin problem~\cite{redner}. 
We need only consider a single $A$ particle and the nearest $C$ particle 
on either side of it, as shown in fig.~\ref{daczero}.
For the probability that the $A$ particle survives, we want the 
probability that all $B$ particles between the $A$ particle and the 
nearest $C$ particles on either side reach the $C$-particles before the 
$A$ particle. The results (after averaging over the distances $L$ and $L'$ 
in Figure \ref{daczero}) will be the same on each side, and independent
so we may solve the backward Fokker-Planck equation (\ref{fokker}) on one
side and simply square the result. We solve Eq.\ (\ref{fokker}) subject to
the boundary conditions $Q(0,t)=0$, $Q(L,t)=1$, where we have an $A$
particle at the origin and a $C$ particle at $x=L$, and the $B$
particle starts at $x$, uniformly distributed in $[0,L]$. The solution is
\begin{eqnarray}\label{mess}
Q(x,t) &=& \frac{x}{L} + \frac{2}{\pi} \sum_{n=1}^{\infty} \frac{1}{n}\, 
\sin\left(\frac{n \pi x}{L} \right) \nonumber \\ 
& & \times \exp\left(- \frac{n^2 \pi^2 D_B t}{L^2}\right).
\end{eqnarray}
Averaging the result over the starting position $x$ gives
\begin{eqnarray}\label{avmess}
Q(t;L) & = & \frac{1}{L} \int_0^L dx\, Q(x,t) \nonumber \\
& = & \frac{1}{2} + \frac{4}{\pi^2} \sum_{\mathrm{odd}\, n} 
\frac{1}{n^2}\,\exp\left(- \frac{n^2\pi^2 D_B t}{L^2}\right).
\end{eqnarray}

We now consider an arbitrary number of $B$ particles in this
interval $[0,L]$. Since the $B$ particles are Poisson distributed,
the probability to have $N$ of them initially in the interval is
\begin{equation*}\label{fish}
p_N = \frac{(\rho_{B0} L)^N}{N!} \rme^{- \rho_{B0} L}.
\end{equation*}
Then the probability that the $A$ survives given $N$ $B$ particles initially 
in $[0,L]$ is, for large $t$, 
\begin{eqnarray}
\nonumber \bar{Q}(t;L) &=& \sum_{N=0}^{\infty} p_N [Q(t; L)]^N 
= \rme^{-\rho_{B0}L[1-Q(t;L)]} \\
\nonumber &\approx & \exp \left( - \frac{\rho_{B0}L}{2} + 
\frac{4 \rho_{B0}L}{\pi^2} \; \rme^{- \frac{\pi^2 D_B t}{L^2} }\right),
\end{eqnarray}
where we have kept only the lowest mode since we are interested in an 
asymptotic large-$t$ result.

Finally, we average over all possible lengths $L$,
weighted by the Poisson distribution for the $C$ particle positions: 
\begin{equation}\label{theint}
Q_A^{(1)}(t) = \rho_C \int_0^\infty \bar{Q}(t;L)\,\rme^{-\rho_C L} dL.
\end{equation}
We simplify by differentiating with respect to $t$ and evaluate the
resulting integral asymptotically for large $t$ using the Laplace
method~\cite{bender}. The result is
\begin{eqnarray} \label{asymp}
\nonumber Q_A^{(1)}(t) & \sim & \frac{1}{1+\frac{\rho_{B0}}{2 \rho_C}}  
\left[ 1+ \frac{8}{(3 \pi)^{1/2}}  (\rho_{B0}^2 D_B t)^{1/2}\right.  \\
\nonumber & & \times \left.  \rme^{-3 \left(\frac{\pi}{2}\right)^{2/3} 
\left(\frac{\rho_C}{\rho_{B0}} + \frac{1}{2}\right)^{2/3} 
(\rho_{B0}^2 D_B t)^{1/3}} \right]
\end{eqnarray}
valid for $\rho_{B0}^2 D_B t \gg 1$. To include the contribution from the 
left side we square this result to obtain, asymptotically,
\begin{eqnarray} \label{final}
\nonumber Q_A(t) & \sim & \frac{1}{\left(1+
\frac{\rho_{B0}}{2 \rho_C}\right)^2}  \left[ 1+ \frac{16}{(3 \pi)^{1/2}}  
(\rho_{B0}^2 D_B t)^{1/2}\right. \\
& & \times \left.  \rme^{-3\left(\frac{\pi}{2}\right)^{2/3} 
\left(\frac{\rho_C}{\rho_{B0}} + \frac{1}{2}\right)^{2/3} 
(\rho_{B0}^2 D_B t)^{1/3}} \right].
\end{eqnarray}

\begin{figure}
\includegraphics[width=\linewidth]{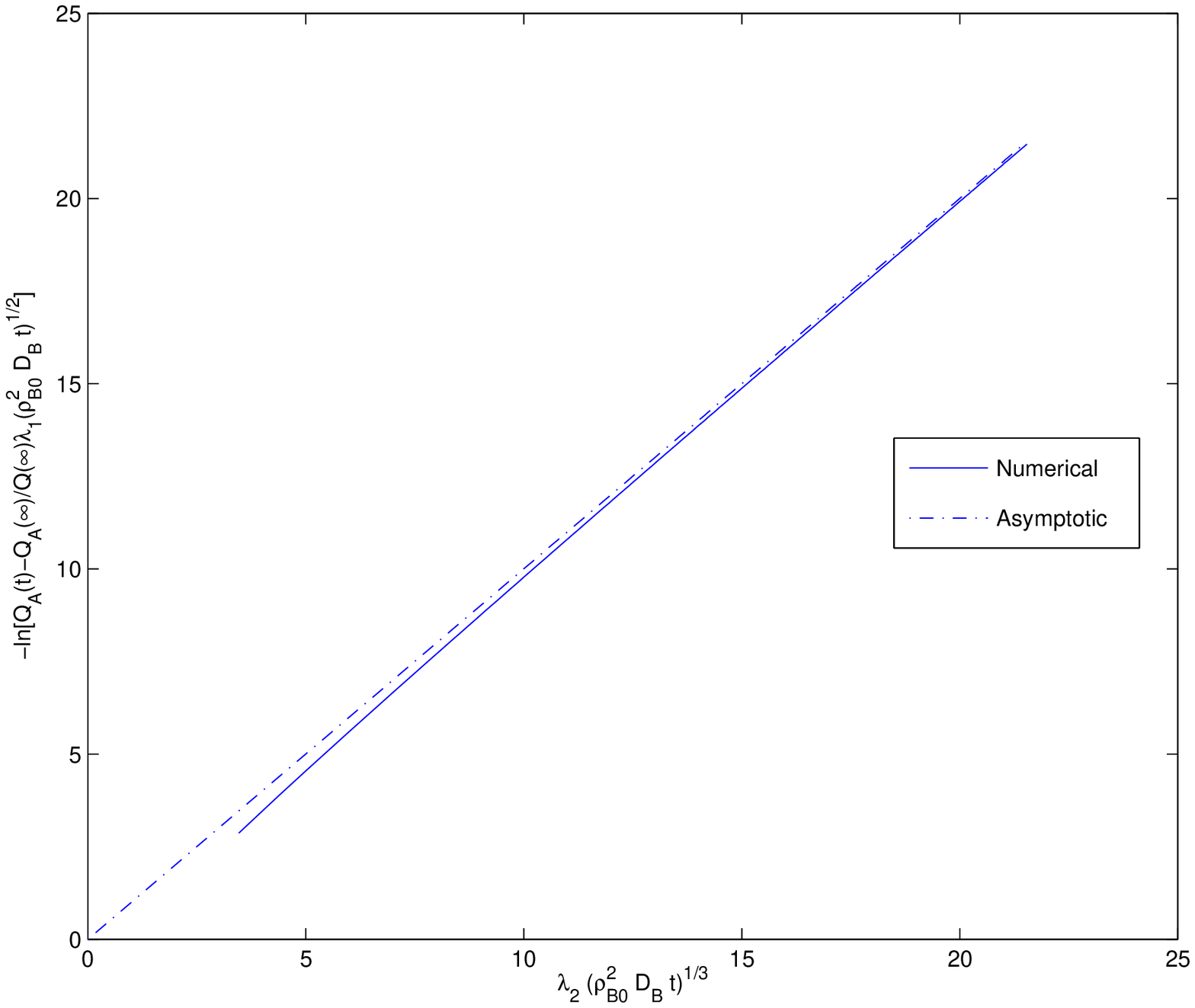}
\caption{\label{robc1} Convergence of the asymptotic solution (\ref{asymp}) 
to the numerical solution of~(\ref{theint}) with $\rho_{B0}/\rho_C = 0.5$. 
The data are presented as a log-linear plot in the form suggested by 
Eq.\ (14). The asymptotic solution has gradient 1.}  

\end{figure}

\begin{figure}
\includegraphics[width=\linewidth]{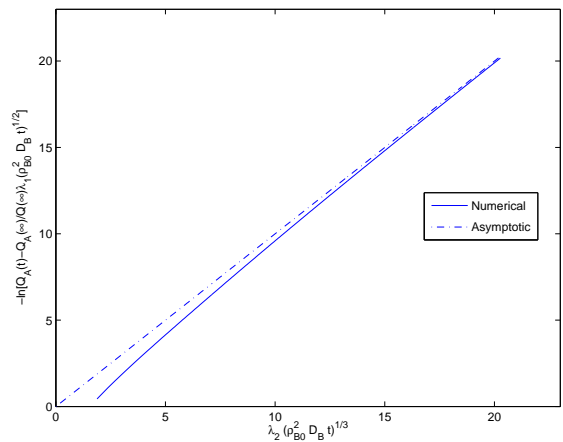}
\caption{\label{robc5} Convergence of the asymptotic solution (\ref{asymp}) 
to the numerical solution of~(\ref{theint}) with $\rho_{B0}/\rho_C = 2$. 
The data are presented as a log-linear plot in the form suggested by 
Eq.\ (14). The asymptotic solution has gradient 1.}
\end{figure}

Note that this result does not require any  condition on the 
ratio $\rho_{B0}/\rho_C$. As a check on the result we evaluate (\ref{theint}) 
numerically using Gauss--Legendre two--point quadrature. We change variables 
to $u=1/(\rho y+1)$, where $\rho = \rho_C + \rho_{B0}/2$, to map the infinite 
range of integration onto the finite interval $[0,1]$. The numerical result 
along with the asymptotic result (\ref{final}) are compared in 
figs.~\ref{robc1} and \ref{robc5} for two values of the ratio 
$\rho_{B0}/\rho_C$, 0.5 and 2 respectively. For ease of comparison 
we first write the result (\ref{final}) in the form 
\begin{equation}
\frac{Q_A(t) - Q_A(\infty)}{Q_A(\infty)\lambda_1(\rho_{B0}^2D_Bt)^{1/2}} 
= \exp[-\lambda_2 (\rho_{B0}^2D_Bt)^{1/3}],
\end{equation}
where $\lambda_1 = 16/\sqrt{3\pi}$ and $\lambda_2 = 3(\pi/2)^{2/3} 
(\rho_C/\rho_{B0} + 1/2)^{2/3}$. The result is plotted in log-linear form 
in figs. \ref{robc1} and \ref{robc5}. 


\section{Discussion and Summary}

The asymptotic  result (\ref{qA}) for  the survival probability  of an
$A$ particle  in the case  $D_A \ll D_B  \ll D_C$ and the  lower bound
(\ref{lowbound})  obtained for  $D_B  =  0$ are  both  subject to  the
condition $\rho_{B0} \ll \rho_C$.  Under these conditions both results
reduce to the same limiting forms as (\ref{sanchez}) at late times:
\begin{equation}
Q_A(\infty)= 1 -
\frac{\rho_{B0}}{\rho_C}\sqrt{\frac{D_{\mathrm{A,B}}}{D_C}},
\end{equation}
correct to first order  in $\rho_{B0}/\rho_C$.  Results (\ref{qA}) and
(\ref{lowbound})  are, however,  of  a more  general  form since  they
indicate   the   nature  of   the   asymptotic   time  dependence   of
$Q_A(t)$.  Under the  condition $\rho_{B0}  \ll \rho_C$  necessary for
these  results  to  be  valid,  the  exponentials  in  (\ref{qA})  and
(\ref{lowbound}) can  be expanded to  first order in  their arguments.

The result (\ref{final}) for the condition $D_A  = D_C = 0$   is still
more  general  since   it  is  valid  for  any   value  of  the  ratio
$\rho_{B0}/\rho_C$.   The exact  infinite-time result  for  this case,
$Q_A(\infty)  = (1+\rho_{B0}/2\rho_C)^{-2}$,  differs from  the result
$(1+\rho_{B0}/\rho_C)^{-1}$  obtained from Eq.\  (\ref{sanchez}) under
the same conditions, although once  more the two results reduce to the
same  limiting   form,  $1-\rho_{B0}/\rho_C$,  to   leading  order  in
$\rho_{B0}/\rho_C$.   These  results  suggest  the  possibility  of  a
systematic expansion in powers of $\rho_{B0}/\rho_C$.

\begin{center}
\begin{small}
{\bf ACKNOWLEDGEMENT}
\end{small}
\end{center}
The work of RS was supported by EPSRC.


\begin{references}

\bibitem{TW}
D. Toussaint and F. Wilczek, J. Chem.\ Phys.\ {\bf 78}, 2642 (1983).

\bibitem{BL}
M. Bramson and J. L. Lebowitz, Phys.\ Rev.\ Lett.\ {\bf 61}, 2397 (1988); 
J. Stat.\ Phys.\ {\bf 62}, 297 (1991).

\bibitem{BB1}
A. J. Bray and R. A Blythe, Phys.\ Rev.\ Lett.\ {\bf 89}, 150601 (2002).

\bibitem{BB2}
R. A. Blythe and A. J. Bray, Phys.\ Rev.\ E {\bf 67}, 041101 (2003). 

\bibitem{AB}
L. Anton and A. J. Bray, J. Phys. A {\bf 37}, 8407 (2004). 

\bibitem{sanchez}
A.D. S\'{a}nchez, E.M. Nicola and H.S. Wio, Phys.\ Rev.\ Lett.\
{\bf 78}, 2244 (1997).

\bibitem{blumen} A. Blumen, G. Zumofen, and J. Klafter, Phys.\
Rev.\ B, {\bf 30}, 5379 (1984).

\bibitem{redner}
S. Redner, {\em A guide to first-passage processes} (CUP, Cambridge, 2001).

\bibitem{odon}
S.J. O'Donoghue and A.J. Bray, Phys.\ Rev.\ E, {\bf 64}, 041105 (2001).

\bibitem{Pascal}
A. J. Bray, S. N. Majumdar, and R. A. Blythe, Phys.\ Rev.\  {\bf 67}, 
060102 (2003); M. Moreau, G. Oshanin, O. B\'enichou, and M. Coppey, 
Phys.\ Rev. E {\bf 67}, 045104 (2003); {\em ibid.} {\bf 69}, 046101 (2004).

\bibitem{weiss}
G.H. Weiss, {\em Aspects and Applications of the Random Walk}
(North--Holland, Amsterdam, 1994).

\bibitem{bender}
C.M. Bender and S.A. Orszag, {\em Advanced Mathematical Methods
for Scientists and Engineers: Asymptotic Methods and Perturbation
Theory} (McGraw--Hill, New York, 1978).

\end{references}
\end{document}